\documentstyle[epsfig]{aipproc}

\input epsf
\let\sec=\section

\newcommand{\etal}{{et~al.}}
\def\japitem#1{\smallskip\noindent\rlap{#1}\hglue\parindent\hangindent\parindent}
\def\enditem{\smallskip\noindent}

\def\japref{\parskip=0pt\par\noindent\hangindent\parindent
    \parskip =2ex plus .5ex minus .1ex}

\def\gs{\mathrel{\lower0.6ex\hbox{$\buildrel {\textstyle >}
 \over {\scriptstyle \sim}$}}}
\def\ls{\mathrel{\lower0.6ex\hbox{$\buildrel {\textstyle <}
 \over {\scriptstyle \sim}$}}}
\newcount\japequationnum
\global\japequationnum=0
\def\bookdisp#1$${\leftline{\hfill{$\displaystyle#1$}
    \global\advance\japequationnum by 1
    \hfill (\the\japequationnum )}$$}
\everydisplay{\bookdisp}
\def\japsub{\rm\scriptscriptstyle}

\def\kms{{\,\rm km\,s^{-1}}}
\def\kmsmpc{{\,\rm km\,s^{-1}Mpc^{-1}}}
\def\hompc{{\,h\,\rm Mpc^{-1}}}
\def\mpcoh{{\,h^{-1}\,\rm Mpc}}

\def\japitem#1{\medskip\noindent\rlap{#1}\hglue 3em\hangindent 3em}

\mathchardef\JAPOmega="700A
\mathchardef\JAPDelta="7001
\mathchardef\JAPLambda="7003
\mathchardef\JAPGamma="7000
\def\japsub{\rm\scriptscriptstyle}
\def\m@th{\mathsurround=0pt }
\def\eqalign#1{\null\,\vcenter{\openup1\jot \m@th
 \ialign{\strut\hfil$\displaystyle{##}$&$\displaystyle{{}##}$\hfil
 \crcr#1\crcr}}\,}

\def\topinsert{\begin{figure}}

\newcommand{\plotter}[2]{\centering \leavevmode \epsfxsize=#2\textwidth \epsfbox{#1}\medskip}
\newcommand{\plottwo}[2]{\centering \leavevmode 
\epsfxsize=0.49\textwidth \epsfbox{#1}
\hglue 1em
\epsfxsize=0.49\textwidth \epsfbox{#2}
\medskip
}

\begin{document}
\title{Measuring large-scale structure with the 2dF Galaxy Redshift Survey}

\author{J.A. Peacock and the 2dFGRS team$^*$}
\address{Institute for Astronomy, University of Edinburgh,\\
Royal Observatory, Edinburgh EH9 3HJ, UK
}


\maketitle

\begin{abstract}
The 2dF Galaxy Redshift Survey is the first to measure more
than 100,000 redshifts. This allows precise measurements of
many of the key statistical measures of galaxy clustering,
in particular redshift-space distortions and the large-scale
power spectrum. 
This paper presents the current 2dFGRS results in these areas.
Redshift-space distortions are detected with a high degree of
significance, confirming the detailed Kaiser distortion from
large-scale infall velocities, and measuring the distortion parameter
$\beta = 0.43 \pm 0.07$.
The power spectrum is measured to $\ls 10\%$ accuracy for
$k>0.02 \hompc$, and is well fitted by a CDM model with
$\Omega_m h =0.20 \pm 0.03$ and a baryon fraction of $0.15\pm 0.07$.

\end{abstract}


\renewcommand{\thefootnote}{\fnsymbol{footnote}}
\footnotetext[1]{{\sl The 2dF Galaxy Redshift Survey team:} Matthew Colless
(ANU), John Peacock (ROE), Carlton M.\ Baugh (Durham), Joss
Bland-Hawthorn (AAO), Terry Bridges (AAO), Russell Cannon (AAO), Shaun
Cole (Durham), Chris Collins (LJMU), Warrick Couch (UNSW), Nicholas
Cross (St Andrews), Gavin Dalton (Oxford), Kathryn Deeley (UNSW),
Roberto De Propris (UNSW), Simon Driver (St Andrews), George Efstathiou
(IoA), Richard S.\ Ellis (Caltech), Carlos S.\ Frenk (Durham), Karl
Glazebrook (JHU), Carole Jackson (ANU), Ofer Lahav (IoA), Ian Lewis
(AAO), Stuart Lumsden (Leeds), Steve Maddox (Nottingham), Darren
Madgwick (IoA), Peder Norberg (Durham), Will Percival (ROE), Bruce
Peterson (ANU), Will Sutherland (ROE), Keith Taylor (Caltech)}
\renewcommand{\thefootnote}{\arabic{footnote}}

\section{Aims and design of the 2dFGRS}

The large-scale structure in the galaxy distribution is widely
seen as one of the most important relics from an early stage of
evolution of the universe.
The 2dF Galaxy Redshift Survey (2dFGRS) was designed  to build on previous
studies of this structure, with the following main aims:
\begin{enumerate}
\item 
To measure the galaxy power spectrum $P(k)$ on scales up to a few
hundred Mpc, bridging the gap between the scales of nonlinear
structure and measurements from the the cosmic microwave background (CMB).
\item
To measure the redshift-space distortion of the large-scale clustering
that results from the peculiar velocity field produced by the mass
distribution. 
\item
To measure higher-order clustering statistics in order to
understand biased galaxy formation, and to test
whether the galaxy distribution on large scales
is a Gaussian random field.
\end{enumerate}

The survey is designed around the 2dF multi-fibre spectrograph on the
Anglo-Australian Telescope, which is capable of observing up to 400
objects simultaneously over a 2~degree diameter field of view. 
Full details of
the instrument and its performance are given in Lewis et~al.\ (2001).
See also {\tt http://www.aao.gov.au/2df/}.

The source catalogue for the survey is a revised and extended version of
the APM galaxy catalogue (Maddox et~al.\ 1990a,b,c). 
The extended version of the APM
catalogue includes over 5~million galaxies down to $b_{\japsub J}=20.5$ in both
north and south Galactic hemispheres over a region of almost
$10^4\, {\rm deg}^2$ (bounded approximately by declination $\delta \leq+3$
and Galactic latitude $b\gs 20$). 
This catalogue is
based on Automated Plate Measuring machine (APM) scans of 390 plates
from the UK Schmidt Telescope (UKST) Southern Sky Survey. The $b_{\japsub J}$
magnitude system for the Southern Sky Survey is defined by the response
of Kodak IIIaJ emulsion in combination with a GG395 filter.
The photometry of the catalogue is calibrated with numerous
CCD sequences and has a precision of approximately 0.2~mag for galaxies
with $b_{\japsub J}=17$--19.5. The star-galaxy separation is as described in
Maddox et~al.\ (1990b), supplemented by visual validation of each galaxy
image.

\begin{figure}
\plotter{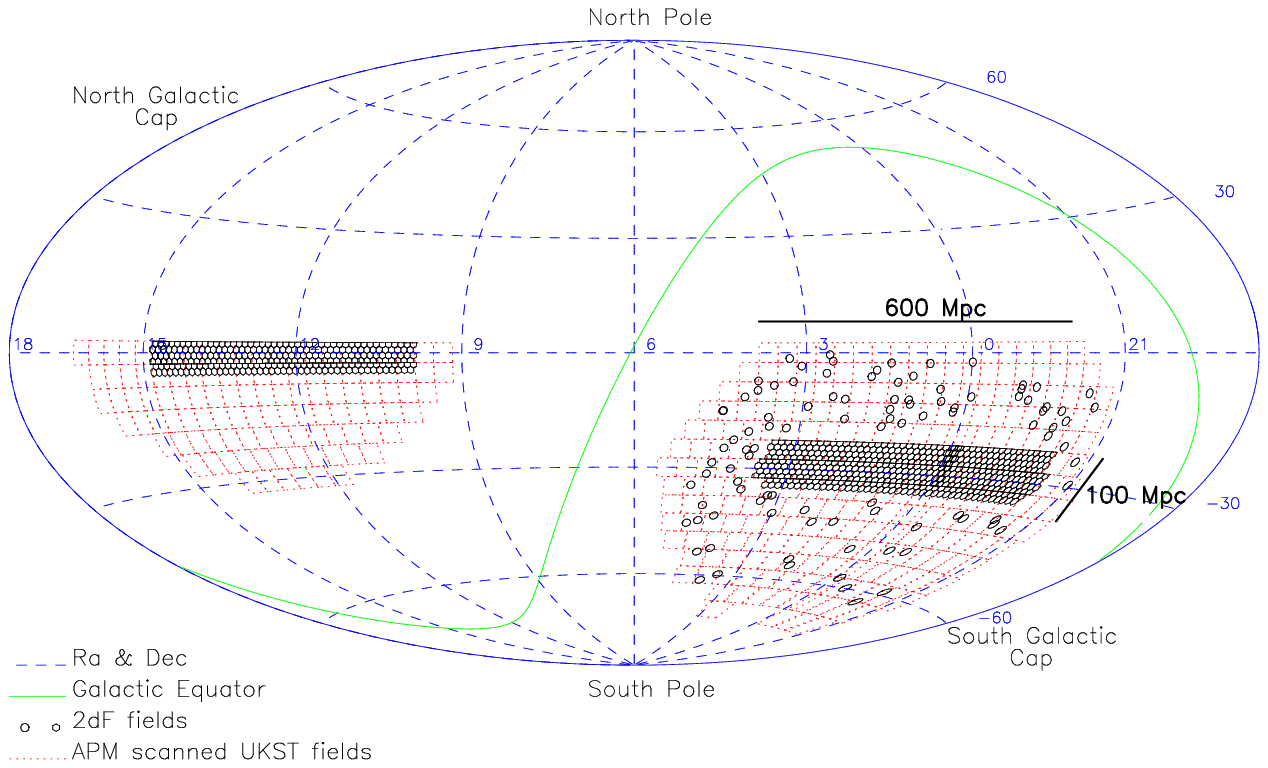}{0.7}
\caption{The 2dFGRS fields (small circles) superimposed on the APM
catalogue area (dotted outlines of Sky Survey plates). There are
approximately 140,000 galaxies in the $75^\circ\times15^\circ$
southern strip centred on the SGP, 70,000 galaxies in the
$75^\circ\times7.5^\circ$ equatorial strip, and 40,000 galaxies in
the 100 randomly-distributed 2dF fields covering the whole area of the
APM catalogue in the south.}
\end{figure}

The survey geometry is shown in Figure~1, and consists of two contiguous
declination strips, plus 100 random 2-degree fields. One strip is in the
southern Galactic hemisphere and covers approximately
75$^\circ$$\times$15$^\circ$ centred close to the SGP at
($\alpha, \delta$)=($01^h$,$-30$); the other strip is in the northern
Galactic hemisphere and covers $75^\circ \times 7.5^\circ$ centred at
($\alpha, \delta$)=($12.5^h$,$+00$). The 100 random fields are spread
uniformly over the 7000~deg$^2$ region of the APM catalogue in the
southern Galactic hemisphere. At the median redshift of the survey
($\bar{z}=0.11$), $100\mpcoh$ subtends about 20~degrees, so the two strips
are $375\mpcoh$ long and have widths of $75\mpcoh$ (south) and $37.5\mpcoh$
(north).

The sample is limited to be brighter than an extinction-corrected
magnitude of $b_{\japsub J}=19.45$ (using the extinction maps of Schlegel et~al.\
1998). This limit gives a good match between the density on the sky of
galaxies and 2dF fibres. Due to clustering, however, the number in a
given field varies considerably. To make efficient use of 2dF, we employ
an adaptive tiling algorithm to cover the survey area with the minimum
number of 2dF fields. With this algorithm we are able to achieve a 93\%
sampling rate with on average fewer than 5\% wasted fibres per field.
Over the whole area of the survey there are in excess of 250,000
galaxies.

\section{Survey Status}

By the end of 2000,
observations had been made of 161,307 targets in 600 fields,
yielding redshifts and identifications for 141,402 galaxies, 7958 stars
and 53 QSOs, at an overall completeness of 93\%. Repeat observations
have been obtained for 10,294 targets. Figure~2 shows the projection of
the galaxies in the northern and southern strips onto $(\alpha,z)$
slices. The main points to note are the level of detail apparent in the
map and the slight variations in density with R.A.\ due to the varying
field coverage along the strips.

\begin{figure}
\plotter{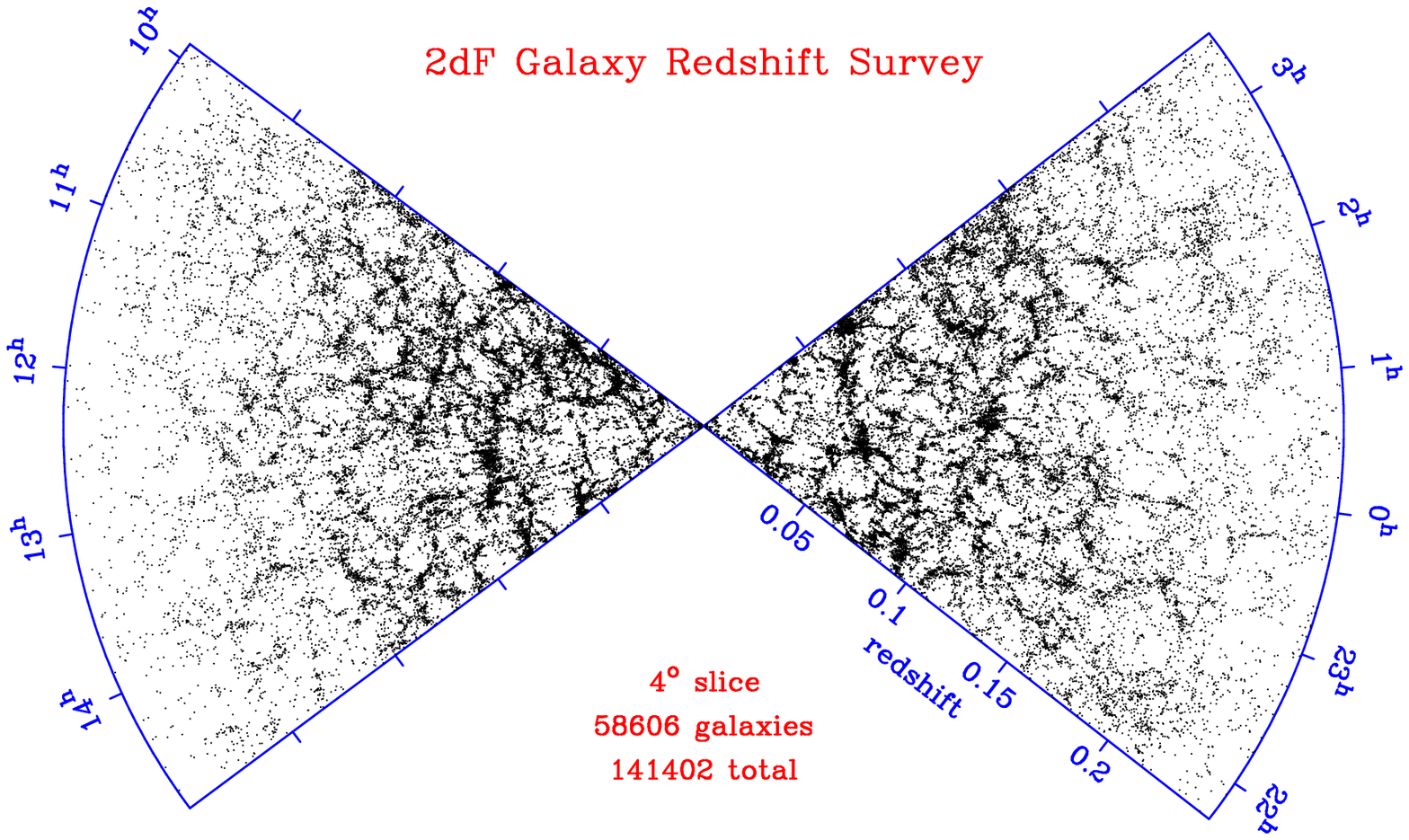}{1.0}
\caption{The distribution of galaxies in part of the 2dFGRS, drawn from 
a total of 141,402 galaxies:
slices $4^\circ$ thick, centred at declination
$-2.5^\circ$ in the NGP and $-27.5^\circ$ in the SGP.
Not all 2dF fields within the slice have been observed
at this stage, hence there are weak variations of the
density of sampling as a function of right ascension.
To minimise such features, the slice thickness increases
to $7.5^\circ$ between right ascension $13.1^h$ and $13.4^h$.
This image reveals a wealth of detail, including
linear supercluster features, often nearly perpendicular
to the line of sight. The interesting question
to settle statistically is whether such transverse features
have been enhanced by infall velocities.}
\end{figure}

The adaptive tiling algorithm is efficient, and yields uniform
sampling in the final survey. However, at this intermediate stage,
missing overlaps mean that the sampling fraction has large fluctuations,
as illustrated in Figure~3.
This variable sampling makes quantification of the large scale
structure more difficult, and limits any analysis requiring relatively
uniform contiguous areas. However, the effective survey `mask' can
be measured precisely enough that it can be allowed for in
low-order analyses of the galaxy distribution.

\begin{figure}
\plotter{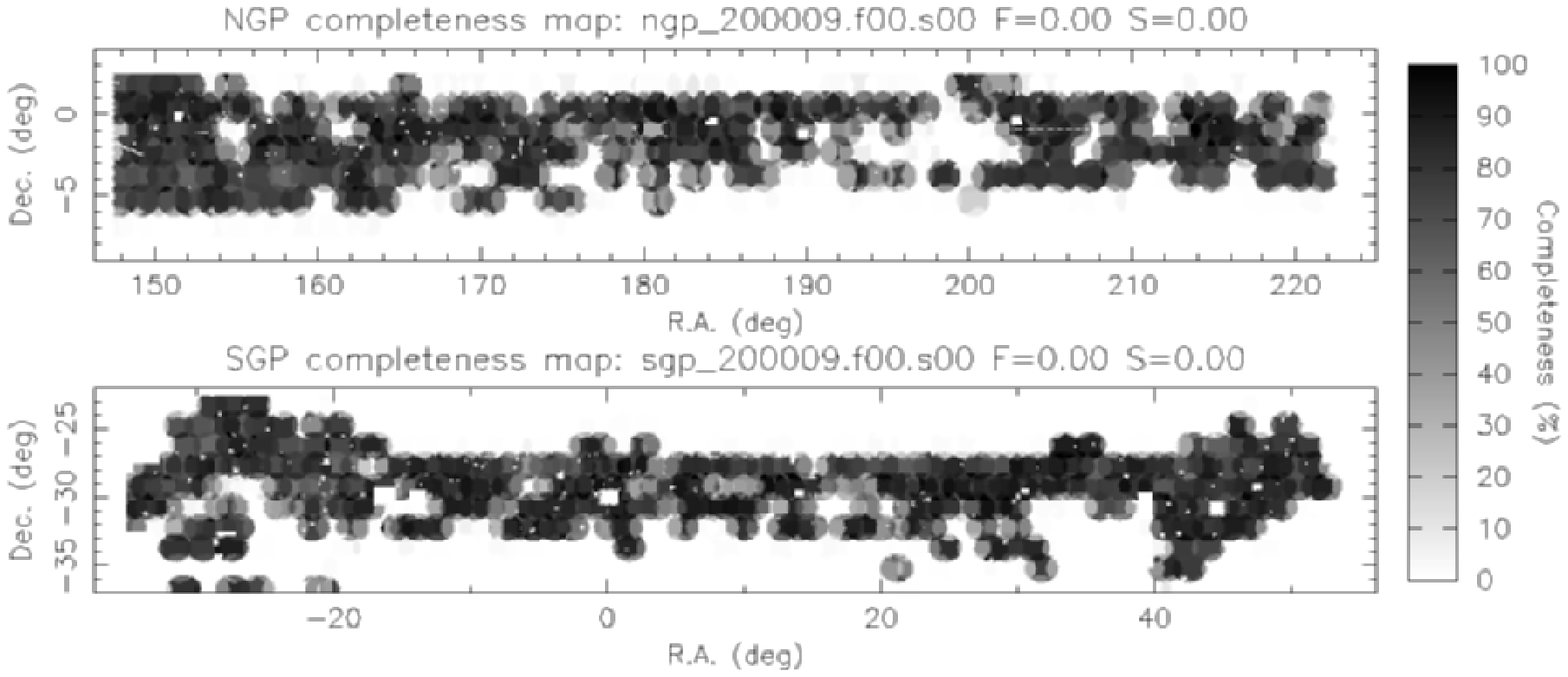}{1.0}
\caption{The completeness as a function of position on the sky. The
circles are individual 2dF fields (`tiles'). Unobserved tiles result in
low completeness in overlap regions. Rectangular holes are omitted
regions around bright stars.}
\end{figure}

\section{Redshift-space correlations}

The simplest statistic for studying clustering in the galaxy
distribution is the the two-point correlation function,
$\xi(\sigma,\pi)$. This measures the excess probability over random of
finding a pair of galaxies with a separation in the plane of the sky
$\sigma$ and a line-of-sight separation $\pi$. Because the radial
separation in redshift space includes the peculiar velocity as well as
the spatial separation, $\xi(\sigma,\pi)$ will be anisotropic. On small
scales the correlation function is extended in the radial direction due
to the large peculiar velocities in non-linear structures such as groups
and clusters -- this is the well-known `Finger-of-God' effect. On large
scales it is compressed in the radial direction due to the coherent
infall of galaxies onto mass concentrations -- the Kaiser effect (Kaiser
1987).

To estimate $\xi(\sigma,\pi)$ we compare the observed count of galaxy
pairs with the count estimated from a random distribution following the
same selection function both on the sky and in redshift
as the observed galaxies. We apply optimal weighting to
minimise the uncertainties due to cosmic variance and Poisson noise.
This is close to equal-volume weighting out to our adopted redshift
limit of $z=0.25$. We have tested our results and found them to be
robust against the uncertainties in both the survey mask and the
weighting procedure.
The redshift-space correlation function for the 2dFGRS 
computed in this way is shown in Figure~4.
The correlation-function results display very clearly
the two signatures of redshift-space distortions discussed
above. The `fingers of God' from small-scale random
velocities are very clear, as indeed has been the case
from the first redshift surveys (e.g. Davis \& Peebles 1983).
However, this is the first time that the 
large-scale flattening from coherent infall has been seen in detail.

\begin{figure}
\plotter{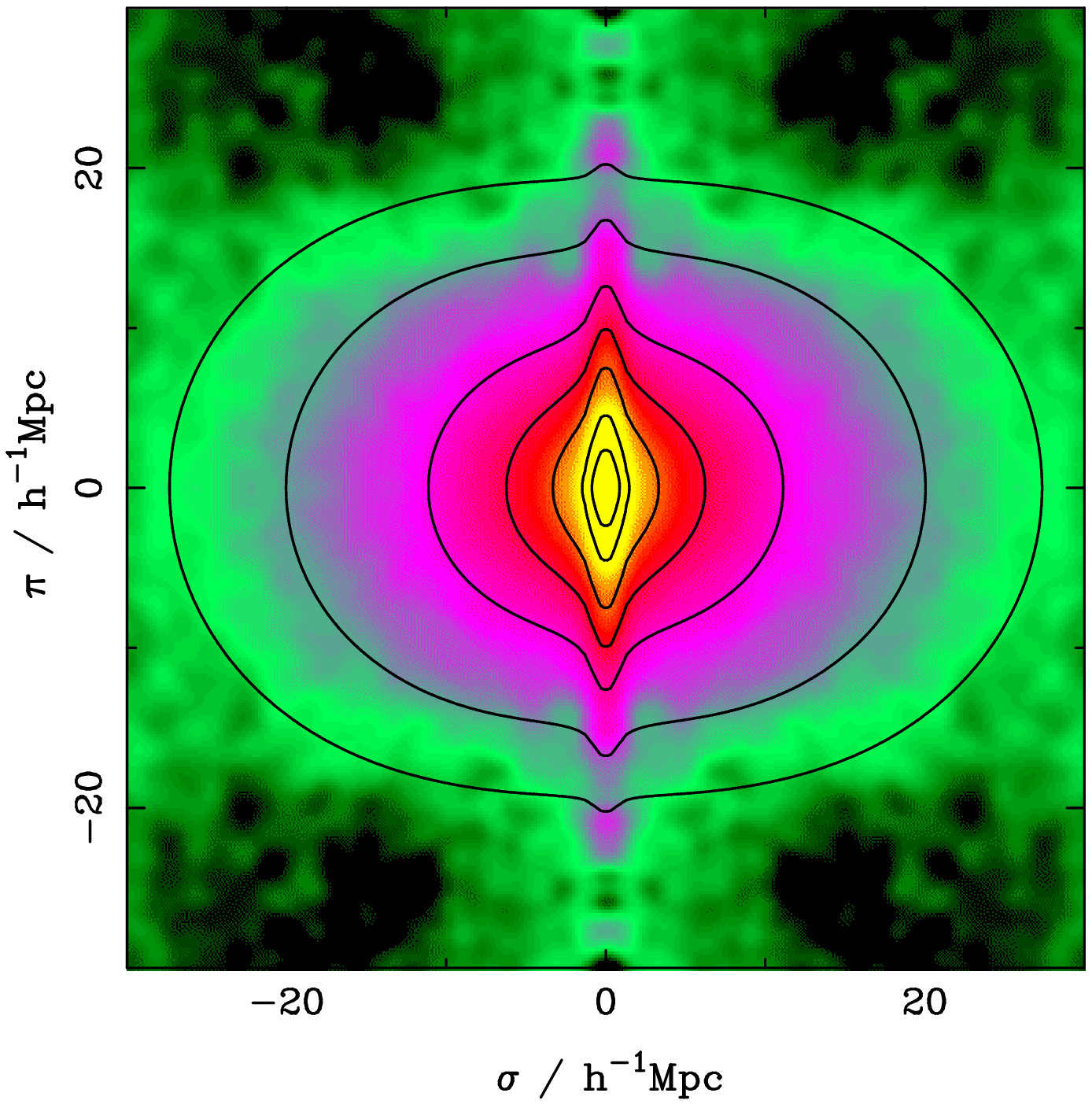}{0.6}
\caption{The galaxy correlation function $\xi(\sigma,\pi)$ as a function
of transverse ($\sigma$) and radial ($\pi$) pair separation is shown as
a greyscale image. It was computed in $0.2\mpcoh$ boxes and then smoothed
with a Gaussian having an rms of $0.5\mpcoh$. The contours are for a model
with $\beta=0.4$ and $\sigma_p=400\kms$, and are plotted at $\xi=10$, 5,
2, 1, 0.5, 0.2 and 0.1.}
\end{figure}

\begin{figure}
\plottwo{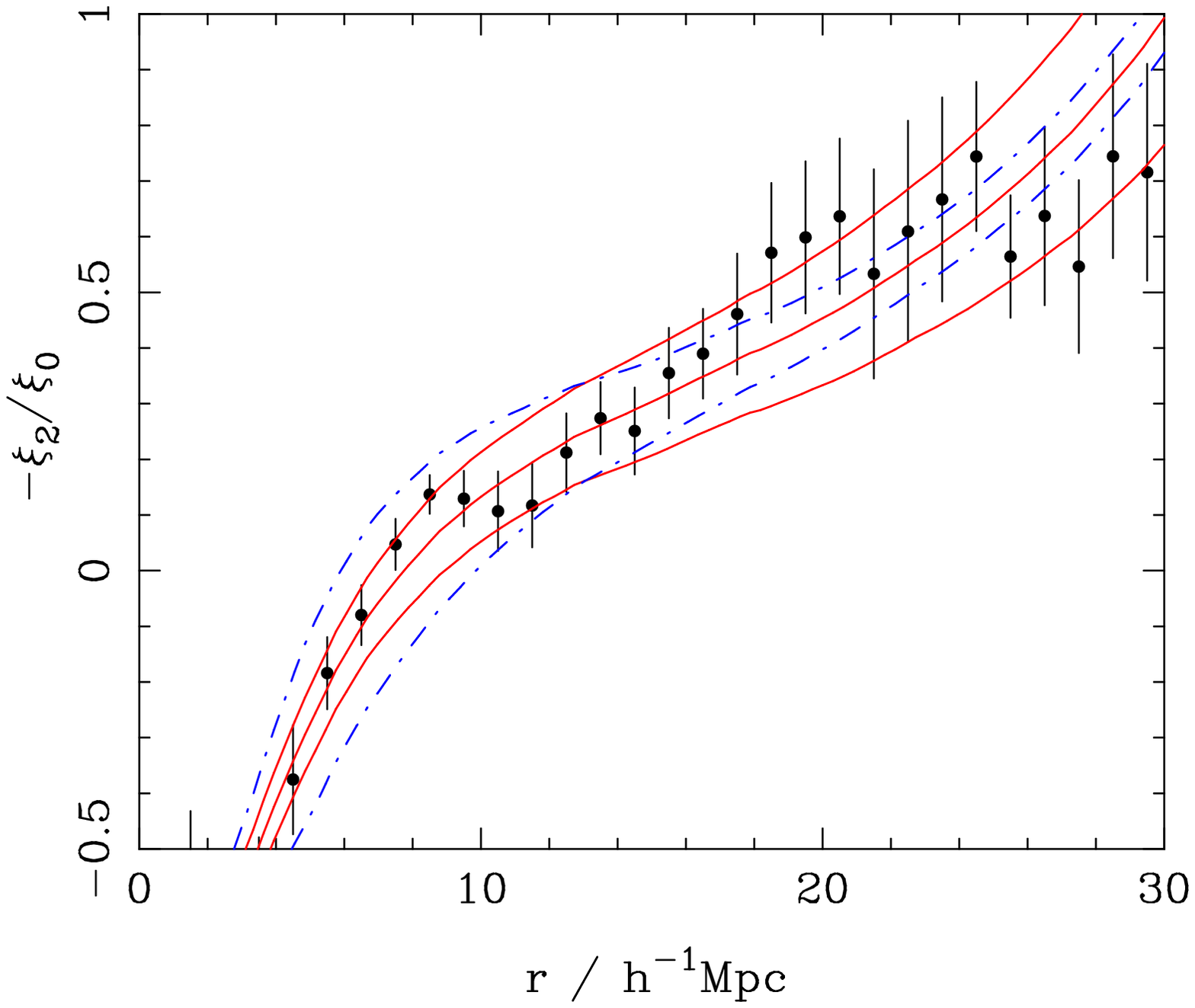}{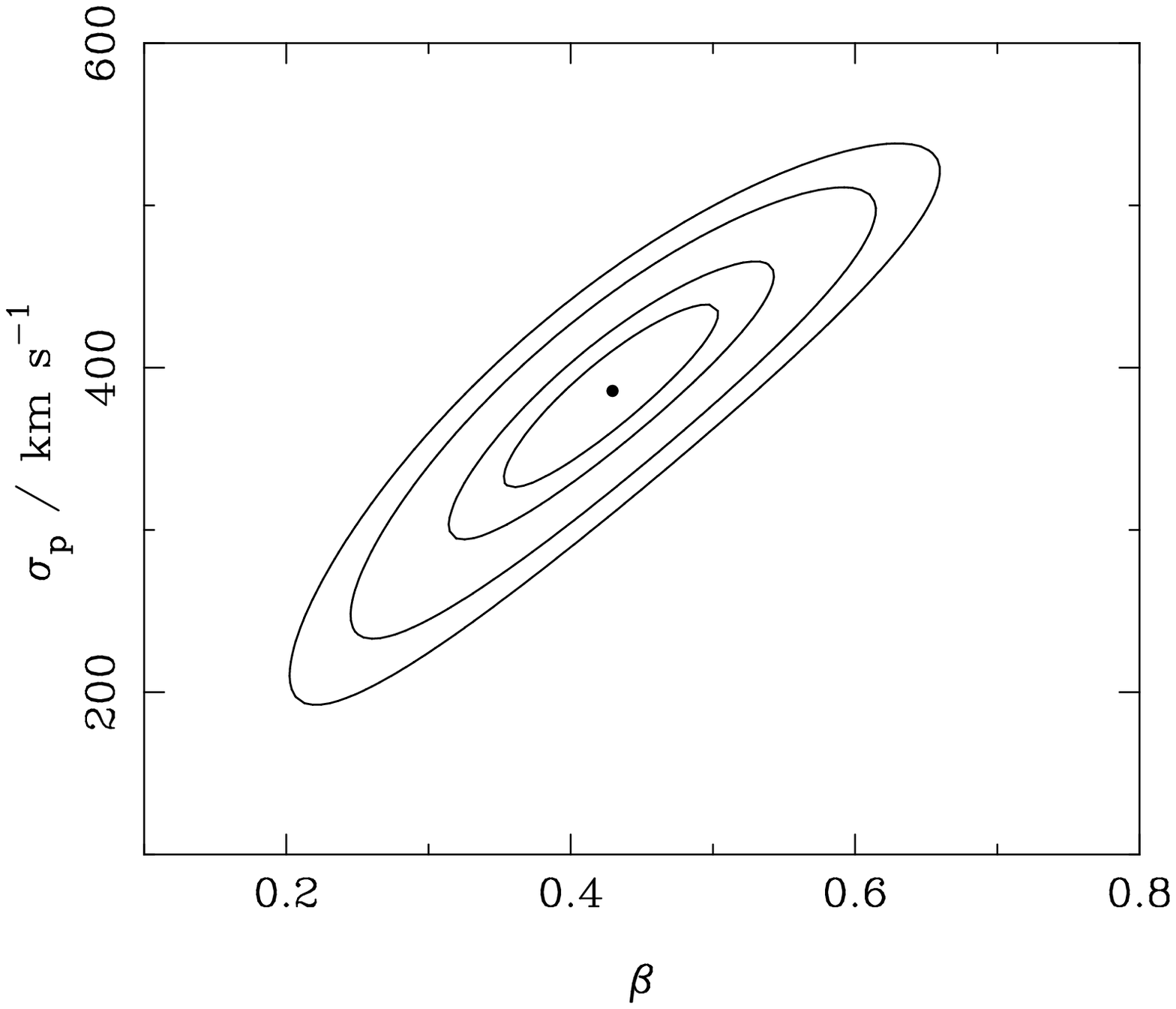}
\caption{(a)~The compression of $\xi(\sigma,\pi)$ as measured by its
quadrupole-to-monopole ratio, plotted as $-\xi_2/\xi_0$. The solid lines
correspond to models with $\sigma_p=400\kms$ and (bottom to top)
$\beta=0.3$,0.4,0.5, while the dot-dash lines correspond to models with
$\beta=0.4$ and (top to bottom) $\sigma_p=300,400,500\kms$.
(b)~Likelihood contours for $\beta$ and $\sigma_p$ from the model fits.
The inner contour is the one-parameter 68\% confidence ellipse; the
outer contours are the two-parameter 68\%, 95\% and 99\% confidence
ellipses. The central dot is the maximum likelihood fit, with
$\beta =0.43$ and $\sigma_p=385\kms$.}
\end{figure}

The degree of large-scale flattening is
determined by the total mass density parameter, $\JAPOmega$, and the
biasing of the galaxy distribution.
On large scales, it should be correct to assume a linear bias model,
so that the redshift-space distortion on large scales depends on
the combination $\beta \equiv \JAPOmega^{0.6}/b$. On these scales, linear
distortions should also be applicable, so we expect to see the
following quadrupole-to-monopole ratio in the correlation function:
$$
\frac{\xi_2}{\xi_0} =
\frac{3+n}{n}\frac{4\beta/3+4\beta^2/7}{1+2\beta/3+\beta^2/5}
$$ 
where $n$ is the power spectrum index of the fluctuations, $\xi \propto
r^{-(3+n)}$. This is modified by the Finger-of-God effect, which is
significant even at large scales and dominant at small scales. 
The effect can be modelled by introducing a parameter
$\sigma_p$, which represents the rms pairwise velocity dispersion of
the galaxies in collapsed structures, $\sigma_p$ (see e.g. Ballinger et al. 1996).
Full details of the fitting procedure are given in Peacock et~al.\ (2001).

Figure~5a shows the variation in $\xi_2/\xi_0$ as a function of scale.
The ratio is positive on small scales where the Finger-of-God effect
dominates, and negative on large scales where the Kaiser effect
dominates. The best-fitting model (considering only the quasi-linear
regime with $8 < r < 25\mpcoh$) has $\beta\simeq 0.4$ and
$\sigma_p \simeq 400\kms$; the likelihood contours are
shown in Figure~5b. Marginalising over $\sigma_p$, the best estimate of
$\beta$ and its 68\% confidence interval is
$$
\beta=0.43\pm0.07
$$
This is the first precise measurement of $\beta$ from redshift-space
distortions; previous studies have shown the effect to exist
(e.g. Hamilton, Tegmark \& Padmanabhan 2000; Taylor et al. 2000; 
Outram, Hoyle \& Shanks 2000), but achieved little more than 3$\sigma$
detections.

\section{Cosmological parameters and the power spectrum}

The detailed measurement of the signature of gravitational
collapse is the first major achievement of the 2dFGRS; we now
consider the quantitative implications of this result.
The first point to consider is that there may be
significant corrections for luminosity effects.
The optimal weighting means that our mean
luminosity is high: it is approximately
1.9 times the characteristic luminosity, $L^*$,  of the overall
galaxy population (Folkes et al. 1999). 
Benoist et al. (1996)  have suggested that the strength of
galaxy clustering increases with luminosity, with
an effective bias that can be fitted by $b/b^* = 0.7 + 0.3(L/L^*)$.
This effect has been controversial (see Loveday et al. 1995), but the 2dFGRS
dataset favours a very similar luminosity dependence.
We therefore expect that $\beta$ for $L^*$ galaxies will exceed our directly measured
figure. Applying a correction using the given formula for $b(L)$,
we deduce $\beta(L=L^*) =  0.54 \pm 0.09.$
Finally, the 2dFGRS has a median redshift of 0.11. With weighting,
the mean redshift in the present analysis is $\bar z =0.17$,
and our measurement should be interpreted as $\beta$ at
that epoch. The extrapolation to $z=0$ is model-dependent,
but probably does not introduce a significant change (Carlberg et al. 2000).

Our measurement of $\JAPOmega^{0.6}/b$ would thus imply 
$\JAPOmega=0.36 \pm 0.10$ if $L^*$ galaxies are unbiased, but it is difficult to justify
such an assumption.
In principle, the details of the clustering pattern in
the nonlinear regime allow the $\JAPOmega - b$ degeneracy to be broken
(Verde et al. 1998), but
for the present it is interesting to use an
independent approach. Observations of CMB
anisotropies can in principle measure 
almost all the cosmological parameters, and 
Jaffe et al. (2000)
obtained the following values
for the densities in collisionless matter ($c$), baryons ($b$),
and vacuum ($v$):
$\JAPOmega_c+\JAPOmega_b+\JAPOmega_v=1.11\pm0.07$,
$\JAPOmega_c h^2=0.14 \pm 0.06$, $\JAPOmega_b h^2 = 0.032\pm 0.005$,
together with a power-spectrum index $n=1.01\pm0.09$.
Our result for $\beta$ gives an independent test of this picture, as follows.

\begin{figure}
\plotter{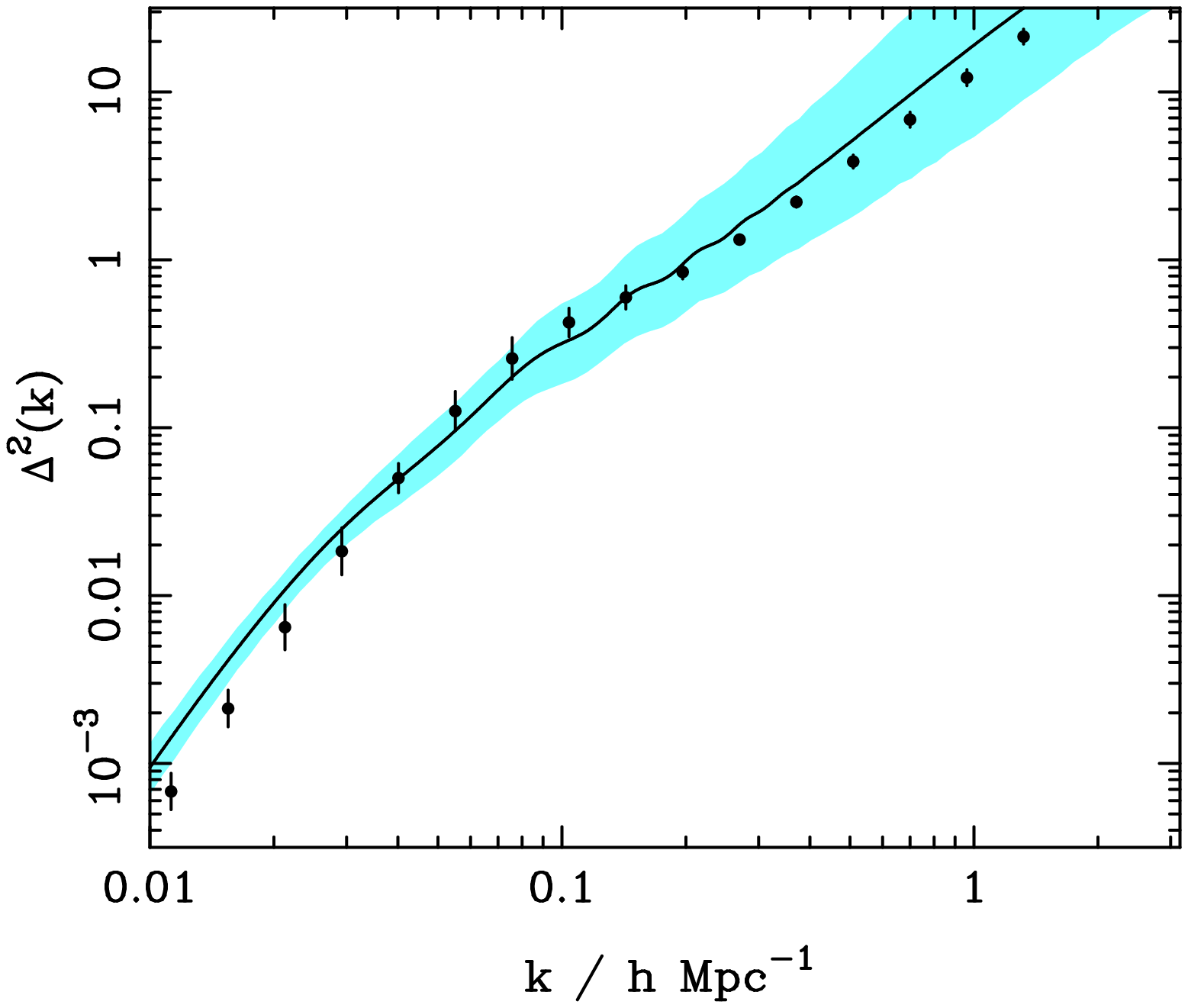}{0.6}
\caption{The dimensionless matter power spectrum at zero
redshift, $\JAPDelta^2(k)$, as predicted from the
allowed range of models that fit the microwave-background
anisotropy data,
plus the assumption that $H_0 = 70 \kmsmpc \pm 10$\%. The solid line shows the
best-fit model from Jaffe et al. (2000) [power-spectrum index $n=1.01$, and
density parameters in baryons, CDM, and vacuum of
respectively 0.065, 0.285, 0.760]. The effects of
nonlinear evolution have been included, according to a revised version
of the procedure of Peacock \& Dodds (1996).
The shaded band shows the $1\sigma$ variation around this
model allowed by the CMB data.
The solid points are the real-space power spectrum measured for APM galaxies.
The clear conclusion is that APM galaxies are consistent
with being essentially unbiased tracers of the mass on large scales.
}
\end{figure}

The only parameter left undetermined by the CMB data is the Hubble constant, $h$.
Recent work (Mould et al. 2000; Freedman et al. 2000) indicates that this is now determined to an
rms accuracy of 10\%,
and we adopt a central value of $h=0.70$. This completes
the cosmological model, requiring a total matter density parameter
$\JAPOmega\equiv \JAPOmega_c+\JAPOmega_b=0.35 \pm 0.14$.
It is then possible to use the parameter limits from the 
CMB to predict a conservative range for 
the mass power spectrum at $z=0$, which is shown in Figure~6.
A remarkable feature of this plot is that the mass power spectrum
appears to be in good agreement with the clustering observed in
the APM survey (Baugh \& Efstathiou 1994). For each model allowed by the CMB, we can predict
both $b$ (from the ratio of galaxy and mass spectra) and also
$\beta$ (since a given CMB model specifies $\JAPOmega$).
Considering the allowed range of models, we then obtain the prediction
$\beta_{\rm\scriptscriptstyle CMB+APM}  = 0.57 \pm 0.17$.
A flux-limited survey such as the APM will have a mean luminosity
close to $L^*$, so the appropriate comparison is with
the 2dFGRS corrected figure of $\beta = 0.54\pm0.09$ for $L^*$ galaxies. These numbers are
in very close agreement.
In the future, the value of $\beta$ will become one of the most
direct ways of confronting large-scale structure with CMB studies.

\sec{The 2dFGRS power spectrum}

Of course, one may question the adoption of the APM
power spectrum, which was deduced by deprojection of angular
clustering. The 3D data of the 2dFGRS should be capable of
improving on this determination, and we have made a first
attempt at doing this, shown in Figure~7.
This power-spectrum estimate uses the FFT-based approach
of Feldman, Kaiser \& Peacock (1994), and needs to be interpreted
with care. Firstly, it is a raw redshift-space estimate, so
that the power beyond $k\simeq 0.2 \hompc$ is severely damped
by fingers of God. On large scales, the power is enhanced, both
by the Kaiser effect and by the luminosity-dependent clustering
discussed above. Finally, the FKP estimator yields the
true power convolved with the window function. This
modifies the power significantly on large scales (roughly
a 20\% correction). We have made an approximate correction for
this in Figure~7 by multiplying by the correction factor appropriate
for a $\JAPGamma=0.25$ CDM spectrum. The precision of the
power measurement appears to be encouragingly high, and the
systematic corrections from the window are well specified.

\begin{figure}
\plotter{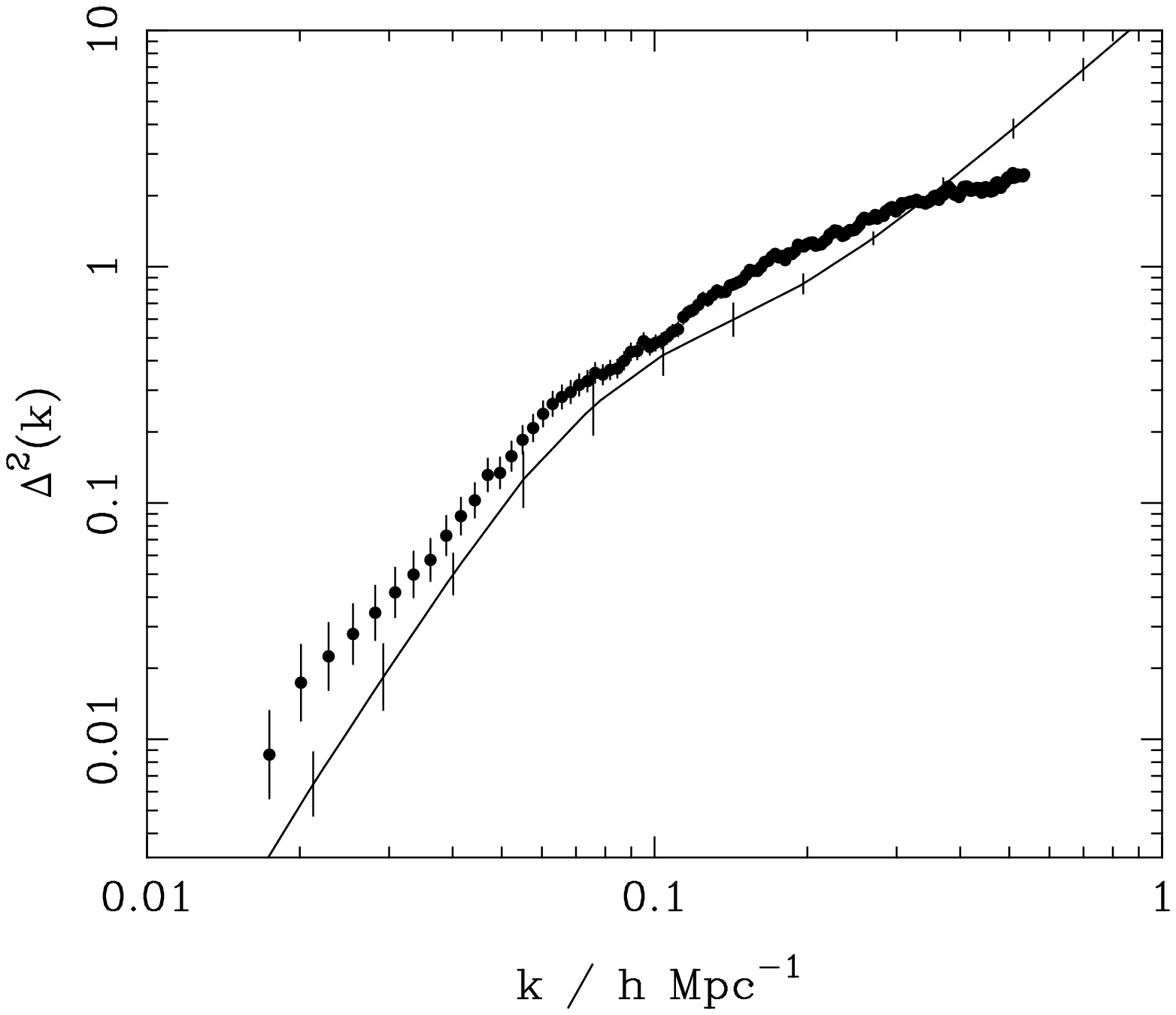}{0.6}
\caption{The 2dFGRS redshift-space power spectrum, 
estimated according to the FKP procedure. The solid points
with error bars show the power estimate. The window
function correlates the results at different $k$ values,
and also distorts the large-scale shape of the power spectrum
An approximate correction for the latter effect has been applied.
The line shows the real-space power spectrum estimated by deprojection
from the APM survey.}
\end{figure}

The next task is to perform a detailed fit of physical
power spectra, taking full account of the window effects.
The hope is that we will obtain not only a more precise measurement
of the overall spectral shape, as parameterized by
$\JAPGamma$, but will be able to move towards more detailed
questions such as the existence of baryonic features in the
matter spectrum (Meiksin, White \& Peacock  1999).
We summarize here results from the first attempt at this analysis
(Percival et al. 2001).

The likelihood of each model has been estimated using a covariance
matrix calculated from Gaussian realisations of linear density fields
for a $\Omega_mh=0.2$, $\Omega_b/\Omega_m=0.15$ CDM power spectrum, for
which $\chi^2_{\rm min}=34.4$, given an expected value of $28$. The
best fit power spectrum parameters are only weakly dependent on this
choice.
The likelihood contours in $\Omega_b/\Omega_m$ versus $\Omega_mh$ for
this fit are shown in Figure~8. At each point in this
surface we have marginalized by integrating the Likelihood surface
over the two free parameters, $h$ and the power spectrum
amplitude. The result is not significantly altered if instead, the
modal, or Maximum Likelihood points in the plane corresponding to
power spectrum amplitude and $h$ were chosen. The likelihood function
is also dependent on the covariance matrix (which should be allowed to
vary with cosmology), although the consistency of result from
covariance matrices calculated for different cosmologies
shows that this dependence is negligibly small.
Assuming a uniform prior for $h$ over a factor of 2 is arguably
over-cautious, and we have therefore added a Gaussian prior $h=0.7\pm
10\%$. This corresponds to multiplying by
the likelihood from external constraints such as the HST key project
(Freedman et al. 2000); this has only a minor effect on the results.

\begin{figure}
\plotter{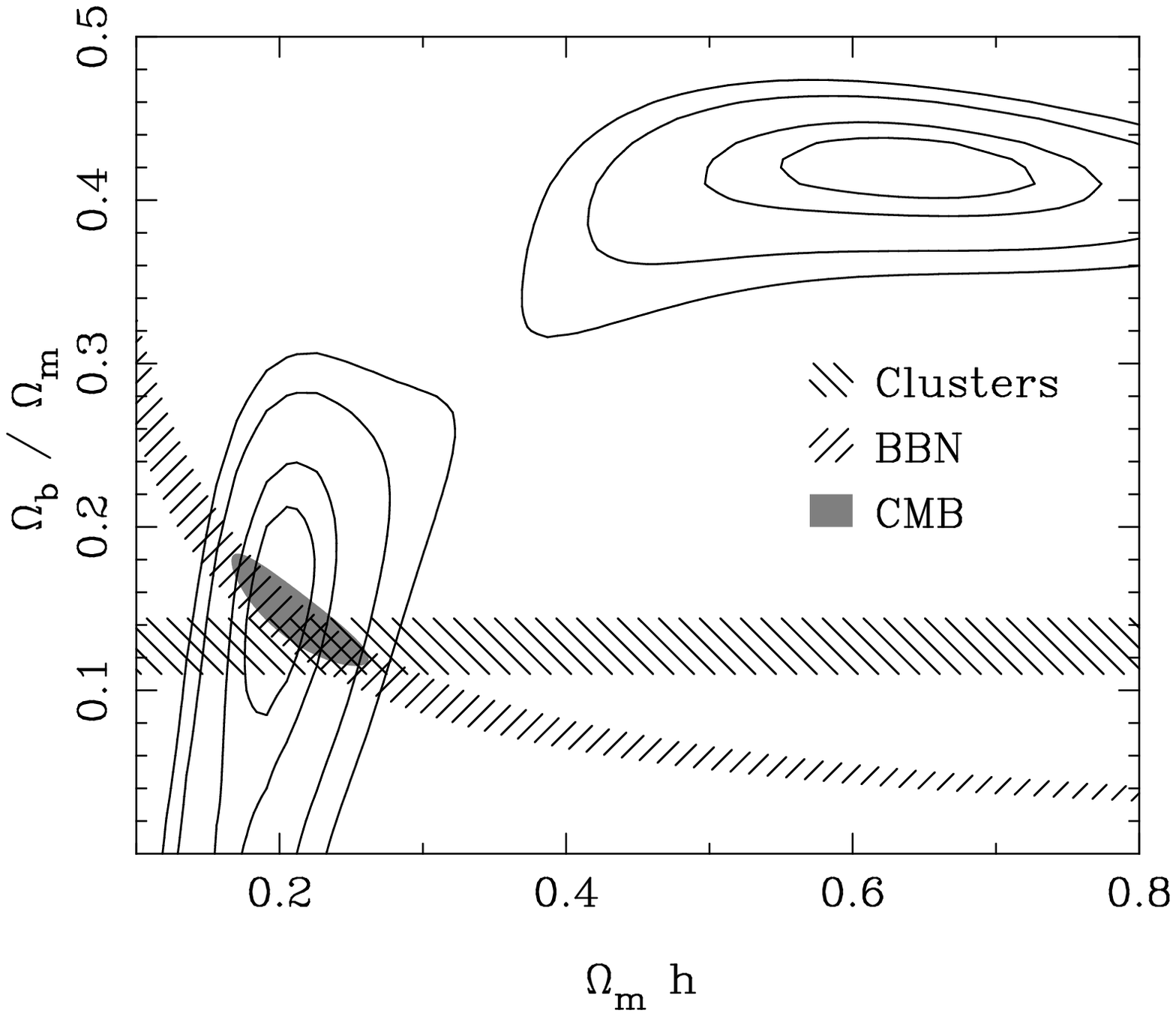}{0.6}
\caption{Likelihood contours for the best-fit linear power spectrum
  over the region $0.02<k<0.15$. The normalization is a free parameter
  to account for the unknown large scale biasing. Contours are plotted
  at the usual positions for one-parameter confidence of 68\%, and
  two-parameter confidence of 68\%, 95\% and 99\% (i.e. $-2\ln({\cal
  L}/{\cal L_{\rm max}}) = 1, 2.3, 6.0, 9.2$). We have marginalized
  over the missing free parameters ($h$ and the power spectrum
  amplitude) by integrating under the Likelihood surface.
  A prior on $h$ of $h=0.7\pm 10\%$ was assumed. 
  This result is compared to estimates from x-ray cluster
  analysis (Evrard 1997), big-bang nucleosynthesis (O'Meara \etal\
  2001) and recent CMB results (Jaffe \etal\ 2000). 
  The CMB results assume that
  $\Omega_bh^2$ and $\Omega_{\rm cdm}h^2$ were independently
  determined from the data.}
\end{figure}

Figure~8 shows that there is a degeneracy between
$\Omega_mh$ and the baryonic fraction $\Omega_b/\Omega_m$. However, there
are two local maxima in the likelihood, one with $\Omega_mh \simeq 0.2$
and $\sim 20\%$ baryons, plus a secondary solution $\Omega_mh \simeq 0.6$
and $\sim 40\%$ baryons. The high-density model can be rejected through a variety
of arguments, and the preferred solution is
$$
  \Omega_m h = 0.20 \pm 0.03; \quad\quad \Omega_b/\Omega_m = 0.15 \pm 0.07.
$$
The 2dFGRS data are compared to the best-fit linear power spectra
convolved with the window function in Figure~9. This
shows where the two branches of solutions come from: the low-density
model fits the overall shape of the spectrum with relatively small
`wiggles', while the solution at $\Omega_m h \simeq 0.6$ provides a
better fit to the bump at $k\simeq 0.065\hompc$, but fits the overall
shape less well.

\begin{figure}
\plotter{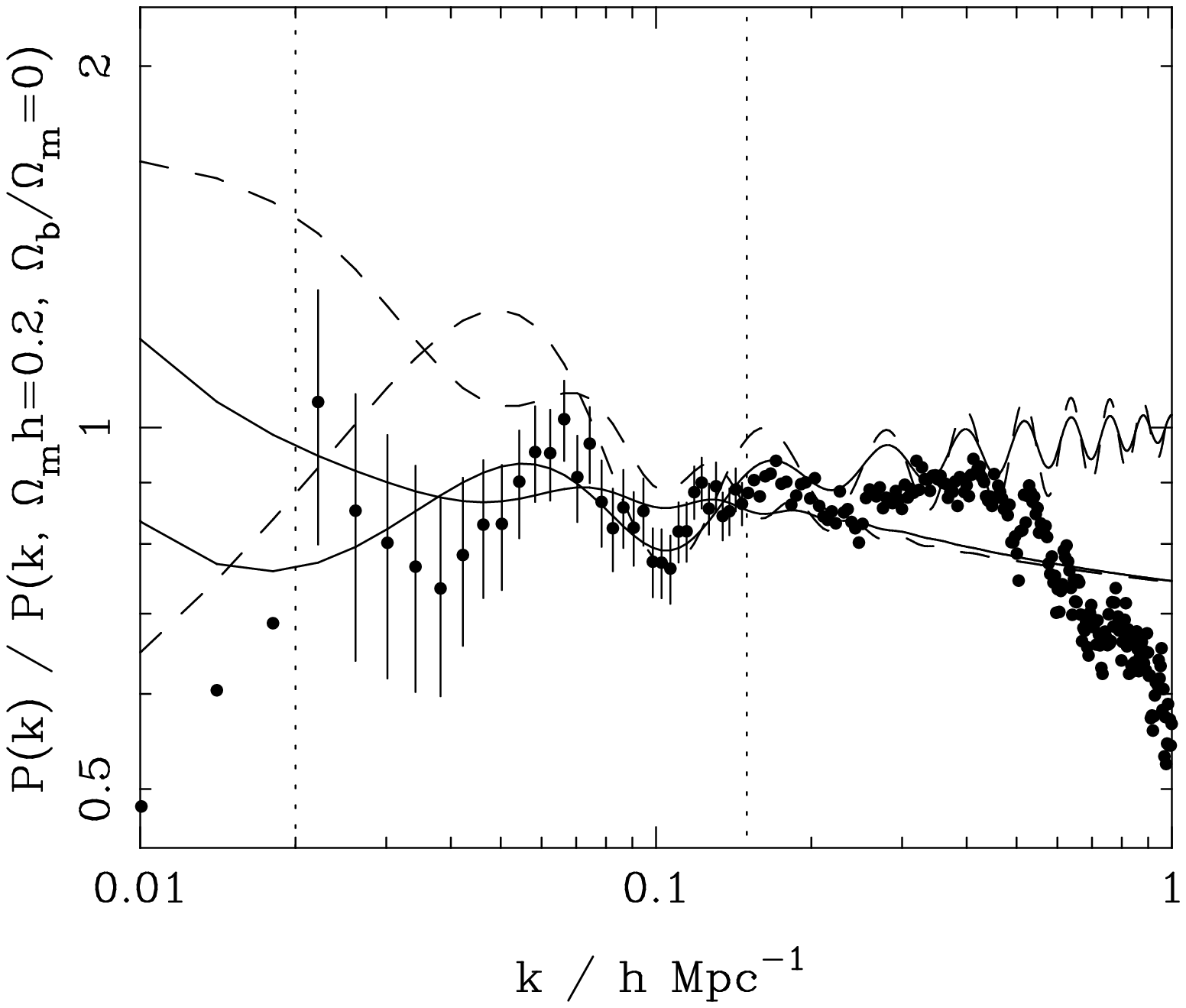}{0.6}
\caption{The 2dFGRS data compared with the two preferred models from
  the Maximum Likelihood fits convolved with the window function
  (solid lines). Error bars show the diagonal elements of the
  covariance matrix, for the fitted data that lie between the dotted
  vertical lines. The unconvolved models are also shown (dashed
  lines). The $\Omega_m h \simeq 0.6$, $\Omega_b/\Omega_m=0.42$,
  $h=0.7$ model has the higher bump at $k\simeq 0.05\hompc$. The
  smoother $\Omega_m h \simeq 0.20$, $\Omega_b/\Omega_m=0.15$, $h=0.7$
  model is a better fit to the data because of the overall shape.}
\end{figure}

Perhaps the main point to emphasize here is that the results are not
greatly sensitive to the assumed tilt of the primordial spectrum. We
have used the CMB results to motivate the choice of $n=1$, but it is
clear that very substantial tilts are required to alter our
conclusions significantly: $n\simeq 0.8$ would be required to turn
zero baryons into the preferred model.

\sec{Conclusions}

The 2dFGRS is now the largest 3D survey of the local
universe, by a factor of over 5 compared to any published survey. 
When it is complete,
we expect to have obtained definitive results on a
number of key issues relating to galaxy clustering.
For details of the current status of the 2dFGRS, see 
{\tt http://www.mso.anu.edu.au/2dFGRS}.
In particular, this site gives details of the 2dFGRS
public release policy, in which we intend to release
approximately the first half of the survey data by
mid-2001, with the complete survey database to be made
public by mid-2003.

At present, the 2dFGRS data allow the galaxy power
spectrum to be measured to high accuracy (10--15\% rms) over about a
decade in scale at $k<0.15\hompc$.  We have carried out a range of
tests for systematics in the analysis and a detailed comparison with
realistic mock samples.  As a result, we are confident that the 2dFGRS
result can be interpreted as giving the shape of the linear-theory
matter power spectrum on these large scales, and that the statistical
errors and covariances between the data points are known.

By fitting our results to the space of CDM models, we have been able
to reach a number of interesting conclusions regarding the matter
content of the universe:

\japitem{(1)}The power spectrum is close in shape to that of a
$\Omega_mh=0.2$ model, to a tolerance of about 20\%.

\japitem{(2)}Nevertheless, there is sufficient structure in the $P(k)$
data that the degeneracy between $\Omega_b/\Omega_m$ and $\Omega_mh$
is weakly broken. The two local likelihood maxima have $(\Omega_mh,
\Omega_b/\Omega_m) \simeq (0.2,0.15)$ and $(0.6,0.4)$ respectively.

\japitem{(3)}Of these two solutions, the preferred one is the
low-density solution. The evidence for detection of baryon
oscillations in the power spectrum is presently modest, with a
likelihood ratio of approximately 3 between the favoured model and the
best zero-baryon model. Conversely, a large baryon fraction can be
very strongly excluded: $\Omega_b/\Omega_m < 0.28$ at 95\% confidence,
provided $\Omega_mh < 0.4$.

\japitem{(4)}These conclusions do not depend strongly on the value of
$h$, but they do depend on the tilt of the primordial spectrum, with
$n\simeq 0.8$ being required to make a zero-baryon model the best fit.

\japitem{(5)}The sensitivity to tilt emphasizes that the baryon signal
comes in good part from the overall shape of the spectrum. Although
the eye is struck by a single sharp `spike' at $k\simeq 0.065\hompc$,
the correlated nature of the errors in the $P(k)$ estimate means that
such features tend not to be significant in isolation. We note that
the convolving effects of the window would require a very substantial
spike in the true power in order to match our data exactly. This is
not possible within the compass of conventional models, and the
conservative conclusion is that the apparent spike is probably
enhanced by correlated noise.  A proper statistical treatment is
essential in such cases.

\enditem
It is interesting to compare these conclusions with other
constraints. According to Jaffe \etal\ (2000), the current CMB data
require $\Omega_m h^2=0.17 \pm 0.06$, $\Omega_b h^2 = 0.032\pm 0.005$,
together with a power-spectrum index of $n=1.01\pm0.09$, on the
assumption of pure scalar fluctuations.  If we take $h=0.7\pm 10\%$,
this gives
\begin{equation}
  \Omega_m h=0.24\pm 0.09;\quad\quad\Omega_b/\Omega_m=0.19\pm 0.07,
\end{equation}
in remarkably good agreement with the estimate from the 2dFGRS
\begin{equation}
  \Omega_m h = 0.20 \pm 0.03; \quad\quad \Omega_b/\Omega_m = 0.15 \pm 0.07.
\end{equation}
Latest estimates of the Deuterium to Hydrogen ratio in QSO spectra
combined with big-bang nucleosynthesis theory predict $\Omega_bh^2 =
0.0205\pm 0.0018$ (O'Meara \etal\ 2001), which disagrees with the CMB
measurement at about the 2$\sigma$ level. The confidence interval
estimated from the 2dFGRS power spectrum overlaps both regions. X-ray
cluster analysis predicts a baryon fraction
$\Omega_b/\Omega_m=0.127\pm0.017$ (Evrard 1997) which is again within
$1\sigma$ of our preferred value.

The above limits are all shown on Figure~9, and paint a
picture of qualitative consistency: it appears that we live in a
universe that has $\Omega_mh\simeq0.2$ with a baryon fraction of
approximately $15\%$. 
It is hard to see how this conclusion can be seriously in error.
Although the CDM model is claimed to have problems in matching
galaxy-scale observations, it clearly works extremely well on large scales.
Any new model that cures the small-scale
problems will have to look very much like $\Omega_m=0.3$ $\Lambda$CDM
on large scales.


\section*{References}

\japref Ballinger W.E., Peacock J.A., Heavens A.F., 1996, MNRAS, 282, 877
\japref Baugh C.M., Efstathiou G., 1994, MNRAS, 267, 323
\japref Benoist C., Maurogordato S., da Costa L.N., Cappi A., Schaeffer R., 1996, ApJ, 472, 452
\japref Carlberg R.G., Yee H.K.C., Morris S.L., Lin H., Hall P.B., Patton D., Sawicki M., Shepherd C.W., 2000, ApJ, 542, 57
\japref Davis M., Peebles, P.J.E., 1983, ApJ, 267, 465
\japref Folkes S.J. et al., 1999, MNRAS, 308, 459
\japref Feldman H.A., Kaiser N., Peacock J.A., 1994, ApJ, 426, 23
\japref Freedman W.L. et al., 2000, astro-ph/0012376
\japref Hamilton A.J.S., Tegmark M., Padmanabhan N., 2000, MNRAS, 317, L23
\japref Jaffe A. et al., 2000, astro-ph/0007333
\japref Kaiser N., 1987, MNRAS, 227, 1
\japref Lewis I., Taylor K., Cannon R.D., Glazebrook K., Bailey J.A., Farrell T.J., Lankshear A., Shortridge K., Smith G.A., Gray P.M., Barton J.R., McCowage C., Parry I.R., Stevenson J., Waller L.G., Whittard J.D., Wilcox J.K., Willis K.C., 2001, MNRAS, submitted
\japref Loveday J., Maddox S.J., Efstathiou G., Peterson B.A., 1995, ApJ, 442, 457
\japref Maddox S.J., Efstathiou G., Sutherland W.J., Loveday J., 1990a, MNRAS, 242, 43{\sc p}
\japref Maddox S.J., Sutherland W.J., Efstathiou G., Loveday J., 1990b, MNRAS, 243, 692
\japref Maddox S.J., Efstathiou G., Sutherland W.J., 1990c, MNRAS, 246, 433
\japref Meiksin A.A., White M., Peacock J.A., 1999, MNRAS, 304, 851
\japref Mould J.R. et al., 2000, ApJ, 529, 786
\japref Outram  P.J., Hoyle F., Shanks T., 2000, astro-ph/0009387
\japref Peacock J.A., Dodds S.J., 1996, MNRAS, 280, L19
\japref Peacock J.A. et al., 2001, Nature, 410, 169
\japref Percival W.J. et al., 2001, MNRAS, submitted
\japref Schlegel D.J., Finkbeiner D.P., Davis M., 1998, ApJ, 500, 525
\japref Taylor A.N., Ballinger W.E., Heavens A.F., Tadros H., 2000, astro-ph/0007048
\japref Verde L., Heavens A.F., Matarrese S., Moscardini L., 1998, MNRAS, 300 747

\end{document}